\documentclass{article}
\usepackage{graphicx}
\begin{document}
\title{The Projector Augmented Wave Method:
ab-initio molecular dynamics with full wave functions}
\author{Peter E. Bl\"ochl$^*$, Clemens J. F\"orst$^{*\dagger}$ 
and Johannes Schimpl$^*$
\\
\\
$^*$Clausthal University of Technology,\\
Institute for Theoretical Physics,\\
 Leibnizstr.10, D-38678 Clausthal-Zellerfeld, Germany
\\
$^\dagger$Vienna University of Technology,\\
Institute for Materials Chemistry,\\
Getreidemarkt 9/165-TC, A-1060 Vienna, Austria}

%
%
%
%
\date{\today} 
\maketitle

\begin{abstract}
A brief introduction to the projector augmented wave method is given
and recent developments are reviewed.  The projector augmented wave
method is an all-electron method for efficient ab-initio molecular
dynamics simulations with full wave functions. It extends and
combines the traditions of existing augmented wave methods and the
pseudopotential approach.  Without sacrificing efficiency, the PAW
method avoids transferability problems of the pseudopotential approach
and it has been valuable to predict properties that depend on the full
wave functions.
\end{abstract}
%
\section{Introduction}

The main goal of electronic structure methods is to solve the
Schr\"odinger equation for the electrons in a molecule or solid, to
evaluate the resulting total energies, forces, response functions and
other quantities of interest. In this paper we review the Projector
augmented wave (PAW) method \cite{PAW94}, an electronic structure
method for ab-initio molecular dynamics with full wave functions. The
main goal of this paper is not to provide a particularly complete or
detailed account of the methodology, but rather to lay out the
underlying ideas. A more rigorous description can be found in the
original paper \cite{PAW94}.

Density functional theory \cite{Kohn64,Kohn65} maps a description for
interacting electrons onto one of non-interacting electrons in an
effective potential.  The remaining one-electron Schr\"odinger
equation still poses substantial numerical difficulties: (1) in the
atomic region near the nucleus, the kinetic energy of the electrons is
large, resulting in rapid oscillations of the wave function that
require fine grids for an accurate numerical representation.  On the
other hand, the large kinetic energy makes the Schr\"odinger equation
stiff, so that a change of the chemical environment has little effect
on the shape of the wave function.  Therefore, the wave function in
the atomic region can be represented well already by a small basis
set.  (2) In the bonding region between the atoms the situation is
opposite. The kinetic energy is small and the wave function is smooth.
However, the wave function is flexible and responds strongly to the
environment. This requires large and nearly complete basis sets.

Combining these different requirements is non-trivial and various
strategies have been developed. 
\begin{itemize}
\item Most appealing to quantum chemists has been the atomic point of
view.  Basis functions are chosen that resemble atomic orbitals. They
exploit that the wave function in the atomic region can be described
by a few basis functions, while the bonding is described by the
overlapping tails of these atomic orbitals. Most techniques in this
class are a compromise of a well adapted basis set with complex matrix
elements on the one hand and on the other hand numerically convenient
basis functions such as Gaussians, where the inadequacies are
compensated by larger basis sets.
\item Pseudopotentials regard an atom as a perturbation of the free
electron gas. The most natural basis functions are plane waves. Plane
waves are complete and well adapted to sufficiently smooth wave
functions. The disadvantage of the large basis sets required is offset
by the extreme simplicity to evaluate matrix elements. Finite plane
wave expansions are, however, absolutely inadequate to describe the
strong oscillations of the wave functions near the nucleus.
In the pseudopotential approach the Pauli repulsion of the core electrons 
is therefore described by an effective potential that expels the
valence electrons from the core region.
The resulting wave functions are smooth and can be represented well by plane waves.  The price to
pay is that all information on the charge density and wave functions
near the nucleus is lost.
\item Augmented wave methods compose their basis functions out of
atom-like partial waves in the atomic regions and a set of functions,
called envelope functions, appropriate for the bonding in between.
Space is divided accordingly into atom-centered spheres, defining the
atomic region, and an interstitial region for the bonds.  The partial
solutions of the different regions, are matched at the interface
between atomic and interstitial regions.
\end{itemize}
The projector augmented wave method is an extension of
augmented wave methods and the pseudopotential approach, which
combines their traditions into a unified electronic structure method.

After describing the underlying ideas of the various methods let us
briefly review the history of augmented wave methods and the
pseudopotential approach. We do not discuss the atomic-orbital based
methods, because our focus is the PAW method and its ancestors.

The augmented wave methods have been introduced in 1937 by
Slater \cite{Slater37} and later modified by Korringa \cite{Korringa47},
Kohn and Rostokker \cite{Kohn54}. They approached the electronic
structure as a scattered electron problem. Consider an electron beam,
represented by a plane wave, traveling through a solid.  It undergoes
multiple scattering at the atoms.  If for some energy, the outgoing
scattered waves interfere destructively, a bound state has been
determined.  This approach can be translated into a basis set method
with energy dependent and potential dependent basis functions.  In
order to make the scattered wave problem tractable, a model potential
had to be chosen: The so-called muffin-tin potential approximates the
potential by a constant in the interstitial region and by a
spherically symmetric potential in the atomic region.

The pseudopotential approach traces back to 1940 when C. Herring
invented the orthogonalized plane wave method \cite{Herring40}. Later,
Phillips \cite{Phillips58} and Antoncik \cite{Antoncik} replaced the
orthogonality condition by an effective potential, that compensates
the electrostatic attraction by the nucleus. In practice, the
potential was modified for example by cutting off the singular potential of the
nucleus at a certain value. This was done with a few parameters that
have been adjusted to reproduce the measured electronic band structure
of the corresponding solid.

Augmented wave and pseudopotential methods reached adulthood in the
1970s: At first O.K.~Andersen \cite{Andersen75} showed that the energy
dependent basis set of Slater's APW method can be mapped onto one with
energy independent basis functions, by linearizing the
partial waves for the atomic regions in energy.  In the original APW
approach the zeros of an energy dependent matrix had to be determined,
which is problematic, if many states lie in a small energy region as
for complex systems.  With the new energy independent basis functions,
however, the problem is reduced to the much simpler generalized
eigenvalue problem, which can be solved using efficient numerical
techniques. Furthermore, the introduction of well defined basis sets
paved the way for full-potential calculations. In that case the
muffin-tin approximation is used solely to define the basis set. The
matrix elements of the Hamiltonian are evaluated with the full
potential.

Hamann, Schl\"uter and Chiang \cite{Hamann79} showed in 1979 how
pseudopotentials can be constructed in such a way, that their
scattering properties are identical to that of an atom to first order
in energy. These first-principles pseudopotentials relieved the
calculations from the restrictions of empirical parameters. Highly
accurate calculations have become possible. 
A main disadvantage of these pseudopotentials has been the large basis set size required
especially for first-row and transition metal atoms.

In 1985 R. Car and M. Parrinello published the ab-initio molecular
dynamics method \cite{Car85}. Simulations of the atomic motion have
become possible on the basis of state-of-the-art electronic structure
methods. Besides making dynamical phenomena and finite temperature
effects accessible to electronic structure calculations, the ab-initio
molecular dynamics method also introduced a radically new way of
thinking into electronic structure methods.  Diagonalization of a
Hamilton matrix has been replaced by classical equations of motion for
the wave function coefficients.  If one applies friction, the system
is quenched to the ground state.  Without friction truly dynamical
simulations of the atomic structure are performed. Electronic wave
functions and atomic positions are treated on equal footing.

The Car-Parrinello method had been implemented first for the pseudopotential
approach. There seemed to be unsurmountable barriers against combining
the new technique with augmented wave methods. The main
problem was related to the potential dependent basis set used sofar:
the Car-Parrinello method requires a well defined and unique total
energy functional of atomic positions and basis set coefficients.
Therefore, it was one of the main goals of the PAW method to introduce energy and potential independent basis sets that were as
accurate and numerically efficient as the previously used augmented
basis sets. Other requirements have been: (1) The method should match
the efficiency of the pseudopotential approach for Car-Parrinello
simulations.  (2) It should become an exact theory when converged and
(3) its convergence should be easily controlled. We believe that these
criteria have been met, which explains why the PAW method becomes
increasingly wide spread today.

We would like to point out that most of these seemingly singular
developments did not come out of the blue, but the ideas seemed to
have evolved in the community. In the case of the PAW method, similar
ideas have been developed by Vanderbilt \cite{Vanderbilt90} in the
context of ultra-soft pseudopotentials. The first dynamical
simulations using a semiempirical electronic structure method have
been performed by Wang and Karplus  \cite{Karplus73} in 1973.  The
first ab-initio pseudopotentials have been published by
Zunger \cite{Zunger78} one year before Hamann, Bachelet and
Schl\"uter \cite{Hamann79}. 

%
\section{Transformation theory}
At the root of the PAW method lies a transformation, that maps the
true wave functions with their complete nodal structure onto auxiliary wave
functions, that are numerically convenient.  We aim for smooth
auxiliary wave functions, which have a rapidly convergent plane wave
expansion. With such a transformation we can expand the auxiliary wave
functions into a convenient basis set, and evaluate all physical
properties after reconstructing the related physical (true) wave functions.

Let us denote the physical one-particle wave functions as
$|\Psi_n\rangle$ and the auxiliary wave functions as
$|\tilde\Psi_n\rangle$. Note that the tilde refers to the
representation of smooth auxiliary wave functions. $n$ is the label
for a one-particle state and contains a band index, a $k$-point and a
spin index. The transformation from the auxiliary to the physical wave
functions is ${\cal T}$.
\begin{eqnarray}
|\Psi_n\rangle={\cal T}|\tilde{\Psi}_n\rangle
\end{eqnarray}

We use here Dirac's Bra and Ket notation. A wave function
$\Psi_n(\mathbf{r})$ corresponds to a ket $|\Psi_n\rangle$, the
complex conjugate wave function $\Psi_n^*(\mathbf{r})$ corresponds to
a bra $\langle\Psi_n|$, and a scalar product $\int d^3r
\Psi_n^*(\mathbf{r})\Psi_m(\mathbf{r})$ is written as
$\langle\Psi_n|\Psi_m\rangle$. Vectors in the 3-d coordinate space are
indicated by boldfaced symbols.

The electronic ground state is determined by minimizing a total energy
functional $E[\Psi_n]$ of the density functional theory. The
one-particle wave functions have to be orthogonal. This constraint is
implemented with the method of Lagrange multipliers. We obtain the
ground state wave functions from the extremum condition for
\begin{eqnarray}
F([\Psi_n],\Lambda_{m,n})=E[\Psi_n]
-\sum_{n,m}[\langle\Psi_n|\Psi_m\rangle-\delta_{n,m}]\Lambda_{n,m}
\end{eqnarray}
with respect to the wave functions and the Lagrange multipliers
 $\Lambda_{n,m}$.  The extremum condition for the wave functions has
 the form
\begin{eqnarray}
 H|\Psi_n\rangle f_n =\sum_m|\Psi_m\rangle\Lambda_{m,n}
\end{eqnarray}
where the $f_n$ are the occupation numbers and
$H=-\frac{\hbar^2}{2m_e}\mathbf{\nabla}^2+v_{\mathrm{eff}}(\mathbf{r})$
is the effective one-particle Hamilton operator.

After a unitary transformation that diagonalizes the matrix of
Lagrange multipliers $\Lambda_{m,n}$, we obtain the Kohn-Sham
equations.
\begin{eqnarray}
H|\Psi_n\rangle=|\Psi_n\rangle\epsilon_n
\end{eqnarray}
The one-particle energies $\epsilon_n$ are the eigenvalues of
$\Lambda_{n,m}\frac{f_n+f_m}{2f_nf_m}$.

Now we express the functional $F$ in terms of our auxiliary wave functions
\begin{eqnarray}
F([{\cal
T}\tilde\Psi_n],\Lambda_{m,n})
= E[{\cal T}\tilde\Psi_n] -\sum_{n,m}[\langle\tilde\Psi_n|{\cal
T}^\dagger{\cal T}|\tilde\Psi_m\rangle -\delta_{n,m}]\Lambda_{n,m}
\end{eqnarray}
The variational principle with respect to the auxiliary wave functions
yields
\begin{eqnarray}
{\cal T}^\dagger H{\cal T}|\tilde\Psi_n\rangle
={\cal T}^\dagger{\cal T}|\tilde\Psi_n\rangle\epsilon_n.
\end{eqnarray}
Again we obtain a Schr\"odinger-like equation, but now the Hamilton
operator has a different form, ${\cal T}^\dagger H{\cal T}$, an overlap
operator ${\cal T}^\dagger{\cal T}$ occurs, and the resulting auxiliary
wave functions are smooth.

When we evaluate physical quantities we need to evaluate expectation
values of an operator $A$, which can be expressed in terms of either the true 
or the auxiliary wave functions.
\begin{eqnarray}
\langle A\rangle&=&\sum_nf_n\langle\Psi_n|A|\Psi_n\rangle
=\sum_nf_n\langle\tilde\Psi_n|{\cal T}^\dagger A{\cal T}|\tilde\Psi_n\rangle
\end{eqnarray}
In the representation of auxiliary wave functions we need to use
transformed operators $\tilde{A}={\cal T}^\dagger A{\cal T}$. 
As it is, this equation only holds for the valence electrons.  
The core electrons are treated differently as will be shown below.

The transformation takes us conceptionally from the world of
pseudopotentials to that of augmented wave methods, which deal with
the full wave functions. We will see that our auxiliary wave
functions, which are simply the plane wave parts of the full wave
functions, translate into the wave functions of the pseudopotential
approach.  In the PAW method the auxiliary wave functions are used to
construct the true wave functions and the total energy functional is
evaluated from the latter.
Thus it provides the missing link between augmented wave methods
and the pseudopotential method, which can be derived as a well-defined
approximation of the PAW method.

In the original paper \cite{PAW94}, the auxiliary wave functions have been termed
pseudo wave functions and the true wave functions have been termed
all-electron wave functions, in order to make the connection more
evident. I avoid this notation here, because it resulted in confusion
in cases, where the correspondence is not clear cut.
%
\section{Transformation operator}
Sofar, we have described how we can determine the auxiliary wave
functions of the ground state and how to obtain physical information
from them.  What is missing, is a definition of the transformation
operator ${\cal T}$.

The operator ${\cal T}$ has to modify the smooth auxiliary wave
function in each atomic region, so that the resulting wave function
has the correct nodal structure.  Therefore, it makes sense to write
the transformation as identity plus a sum of atomic contributions
${\cal S}_R$
\begin{eqnarray}
{\cal T}=1+\sum_R{\cal S}_R.
\end{eqnarray}
For every atom, ${\cal S}_R$ adds the difference between the true and
the auxiliary wave function. The index $R$ is a label for an atomic
site.

The local terms ${\cal S}_R$ are defined in terms of solutions
$|\phi_{i}\rangle$ of the Schr\"odinger equation for the isolated
atoms.  This set of partial waves $|\phi_{i}\rangle$ will serve as a
basis set so that, near the nucleus, all relevant valence wave functions can be expressed
as superposition of the partial waves with yet unknown coefficients.
\begin{eqnarray}
\Psi(\mathbf{r})=\sum_{i\in R}\phi_{i}(\mathbf{r}) c_i\quad {\rm
for}\quad |\mathbf{r}-\mathbf{R}_R|<r_{c,R}
\end{eqnarray}
The index $i$ refers to a site index $R$, the angular momentum indices
$(\ell,m)$ and an additional index that differentiates partial waves
with same angular momentum quantum numbers on the same site. With
$i\in R$ we indicate those partial waves that belong to site
$R$. $\mathbf{R}_R$ is the position of the nucleus of site $R$.

Note that the partial waves are not necessarily bound states and
are therefore not normalizable, unless we truncate them beyond a
certain radius $r_{c,R}$. The PAW method is formulated such that the
final results do not depend on the location where the partial waves
are truncated, as long as this is not done too close to the nucleus.

Since the core wave functions do not spread out into the neighboring
atoms, we will treat them differently. Currently we use the
frozen-core approximation so that density and energy of the core
electrons are identical to those of the corresponding isolated
atoms. The transformation ${\cal T}$ shall produce only wave functions
orthogonal to the core electrons, while the core electrons are treated
separately.  Therefore, the set of atomic partial waves
$|\phi_i\rangle$ includes only valence states that are orthogonal
to the core wave functions of the atom.

For each of the partial waves we choose an auxiliary partial wave
$|\tilde\phi_i\rangle$. The identity
\begin{eqnarray}
|\phi_i\rangle&=&(1+S_R)|\tilde\phi_i\rangle
\quad\mathrm{for}\quad i\in R 
\nonumber\\
S_R|\tilde\phi_i\rangle&=&|\phi_i\rangle-|\tilde\phi_i\rangle
\label{eq:sr1}
\end{eqnarray}
defines the local contribution ${\cal S}_R$ to the transformation
operator. Since $1+{\cal S}_R$ shall change the wave function only
locally, we require that the partial waves $|\phi_i\rangle$ and their
auxiliary counter parts $|\tilde\phi_i\rangle $ are pairwise identical
beyond a certain radius $r_c$.
\begin{eqnarray}
\phi_i(r)=\tilde\phi_i(r)
\quad\mathrm{for}\quad i\in R\quad\mathrm{and}\quad|\mathbf{r}-\mathbf{R}_R|
>r_{c,R}
\label{eq:equaloutside}
\end{eqnarray}

\begin{figure}[ht]
\centering
\includegraphics[width=1.5cm,angle=-90,clip=true]{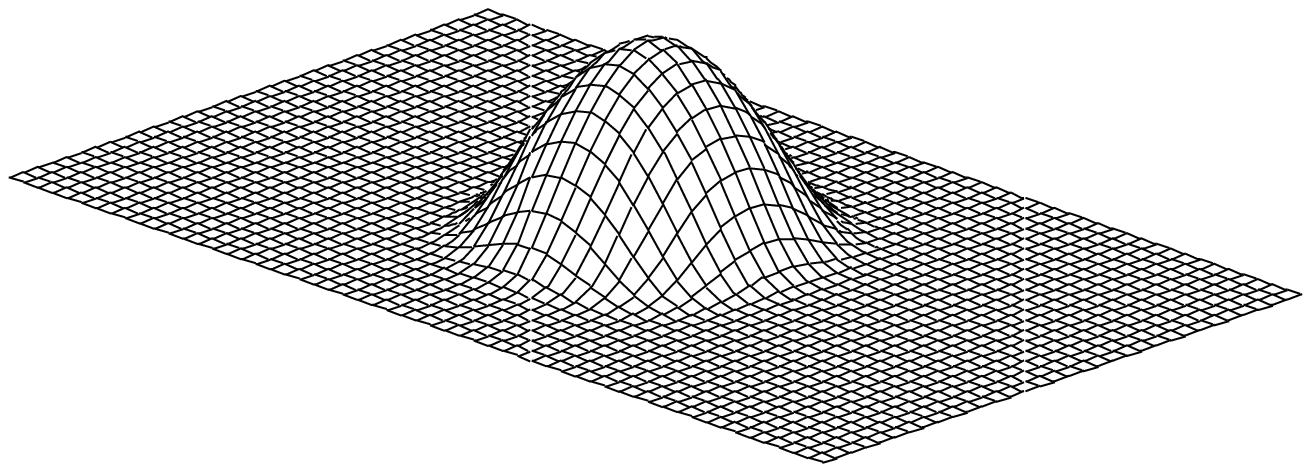}\includegraphics[width=1.5cm,angle=-90,clip=true]{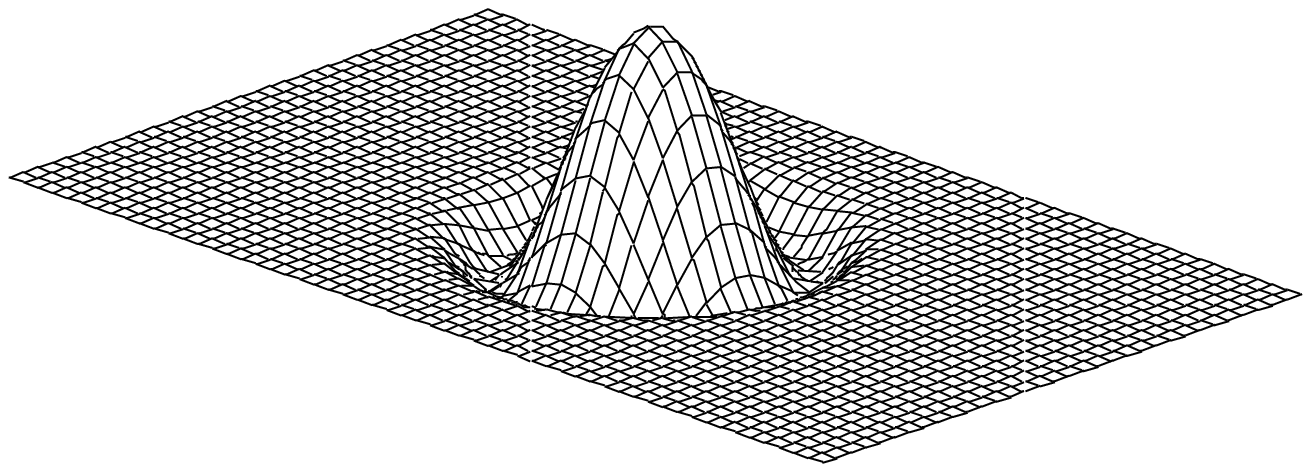}\\
\includegraphics[width=1.5cm,angle=-90,clip=true]{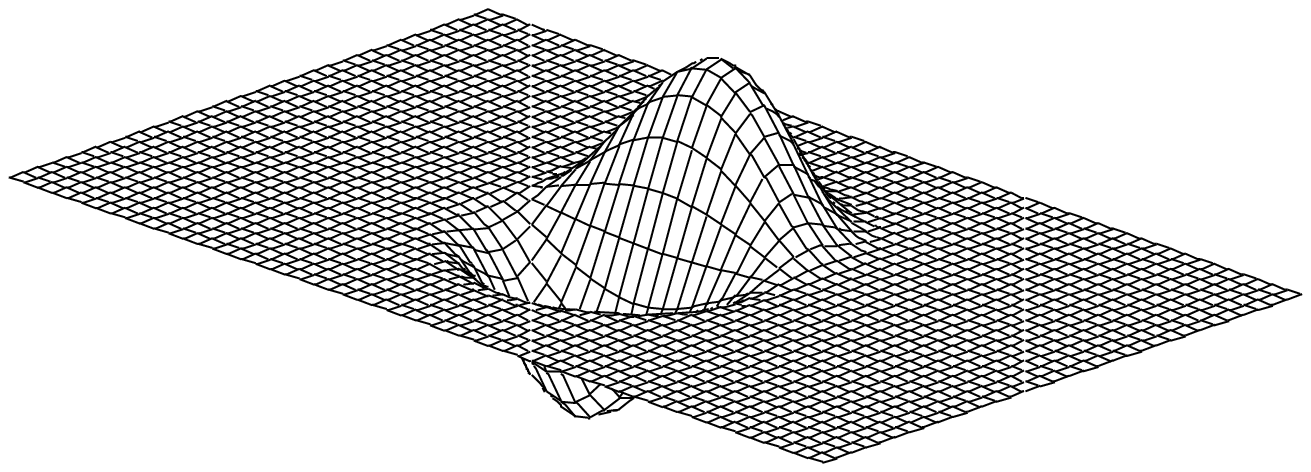}\includegraphics[width=1.5cm,angle=-90,clip=true]{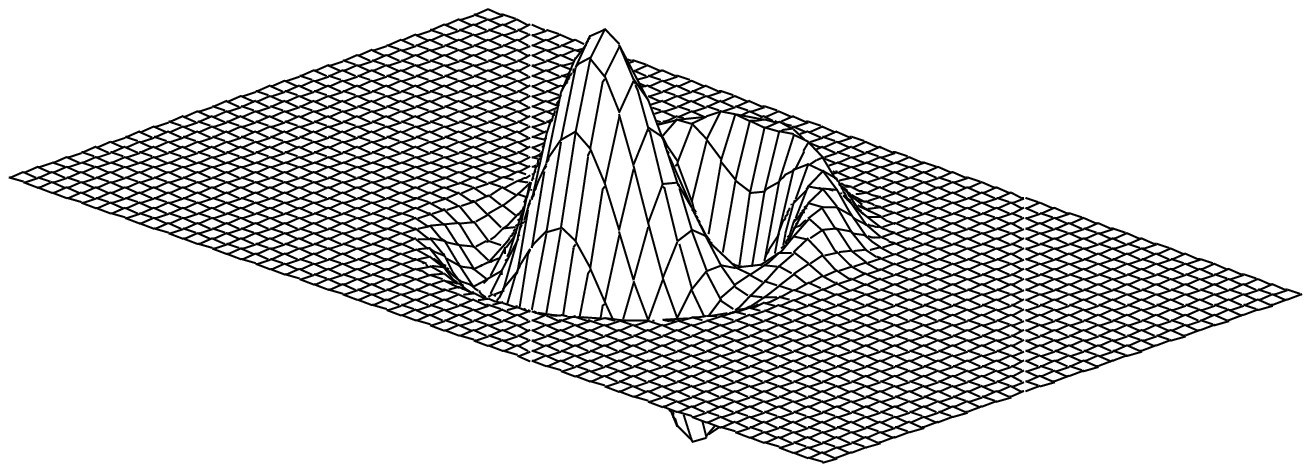}\\
\includegraphics[width=1.5cm,angle=-90,clip=true]{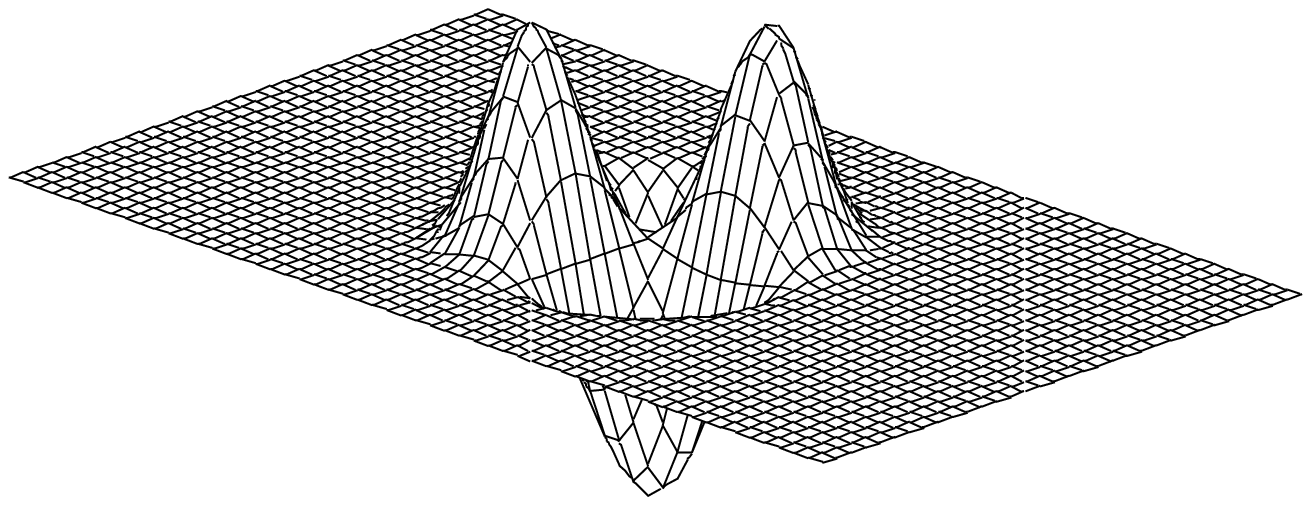}
\caption{Top: projector functions of the Cl atom for two s-type
partial waves, middle: p-type, bottom: d-type.}
\label{fig:2}
\end{figure}

In order to be able to apply the transformation operator to an
arbitrary auxiliary wave function, we need to be able to expand the
auxiliary wave function locally into the auxiliary partial waves.
\begin{eqnarray}
\tilde\Psi(\mathbf{r})&=&\sum_{i\in R} \tilde\phi_i(\mathbf{r})
\langle\tilde{p}_i|\tilde\Psi\rangle  
\quad{\rm for}\quad |{\bf r}-{\bf R}_R|<r_{c,R}
\label{eq:ps1center}
\end{eqnarray}
which defines the projector functions $|\tilde{p}_i\rangle$.  The projector
functions probe the local character of the auxiliary wave function in the
atomic region.  Examples of projector functions are shown in Fig.~\ref{fig:2}.
From Eq.~\ref{eq:ps1center} we can derive
$\sum_i|\tilde\phi_i\rangle\langle\tilde{p}_i|=1$, which is valid within $r_c$.
It can be shown by insertion, that the identity Eq.~\ref{eq:ps1center} holds
for any auxiliary wave function $|\tilde\Psi\rangle$ that can be expanded
locally into auxiliary partial waves $|\tilde\phi_i\rangle$, if
\begin{equation}
\langle\tilde{p}_i|\tilde\phi_j\rangle=\delta_{i,j} 
\quad\mathrm{for}\quad i,j\in R
\end{equation}
Note that neither the projector functions nor the partial waves need
to be orthogonal among themselves.  

By combining Eq.~\ref{eq:sr1} and Eq.~\ref{eq:ps1center}, we can apply
$S_R$ to any auxiliary wave function.
\begin{eqnarray}
S_R|\tilde\Psi\rangle
&=&\sum_{i\in R} S_R |\tilde\phi_i\rangle\langle\tilde{p}_i|\tilde\Psi\rangle
=\sum_{i\in R} \Bigl(|\phi_i\rangle-|\tilde\phi_i\rangle\Bigr)
\langle\tilde{p}_i|\tilde\Psi\rangle
\end{eqnarray}

Hence the transformation operator is
\begin{eqnarray}
{\cal T}=1+\sum_i\Bigl(|\phi_i\rangle-|\tilde\phi_i\rangle\Bigr)
\langle\tilde{p}_i|
\label{eq:transf}
\end{eqnarray}
where the sum runs over all partial waves of all atoms.  The true wave
function can be expressed as
\begin{eqnarray}
|\Psi\rangle=|\tilde\Psi\rangle
+\sum_i\Bigl(|\phi_i\rangle-|\tilde\phi_i\rangle\Bigr)
\langle\tilde{p}_i|\tilde\Psi\rangle
=|\tilde\Psi_n\rangle
+\sum_R\Bigl( |\Psi^1_R\rangle-|\tilde\Psi^1_R\rangle\Bigr)
\label{eq:aewave}
\end{eqnarray}
with
\begin{eqnarray}
|\Psi^1_R\rangle&=&\sum_{i\in R}|\phi_i\rangle
\langle\tilde{p}_i|\tilde\Psi\rangle
\\
|\tilde\Psi^1_R\rangle&=&\sum_{i\in R}|\tilde\phi_i\rangle
\langle\tilde{p}_i|\tilde\Psi\rangle
\end{eqnarray}

\begin{figure}[h]
\centering
\includegraphics[height=6cm,angle=-90,clip=true]{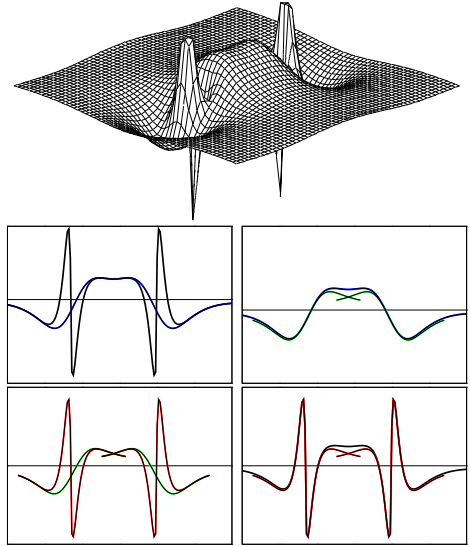}\\
\includegraphics[width=3cm,clip=true]{ae-ps_new.eps}
\includegraphics[width=3cm,clip=true]{ps-ps1_new.eps}\\
\includegraphics[width=3cm,clip=true]{ae1-ps1_new.eps}
\includegraphics[width=3cm,clip=true]{ae-ae1_new.eps}
\caption{Bonding p-$\sigma$ orbital of the Cl$_2$ molecule and its
decomposition of the wave function into auxiliary wave function and the
two one-center expansions.  Top-left: True and auxiliary wave
function; top-right: auxiliary wave function and its partial wave
expansion; bottom-left: the two partial wave expansions; bottom-right:
true wave function and its partial wave expansion.}
\label{fig:1}
\end{figure}

In Fig.~\ref{fig:1} the decomposition of Eq.~\ref{eq:aewave} is shown
for the example of the bonding p-$\sigma$ state of the Cl$_2$
molecule.

To understand the expression for the true wave function,
Eq.~\ref{eq:aewave}, let us concentrate on different regions in
space. (1) Far from the atoms, the partial waves are, according to
Eq.~\ref{eq:equaloutside}, pairwise identical so that the auxiliary
wave function is identical to the true wave function
$\Psi(\mathbf{r})=\tilde\Psi(\mathbf{r})$. (2) Close to an atom,
however, the true wave function $\Psi(\mathbf{r})
=\Psi^1_R(\mathbf{r})$ is built up from partial waves that contain the
proper nodal structure, because the auxiliary wave function and its
partial wave expansion are equal according to Eq.~\ref{eq:ps1center}.

In practice the partial wave expansions are truncated.  Therefore, the
identity of Eq.~\ref{eq:ps1center} does not hold strictly.  As a
result the plane waves also contribute to the true wave function
inside the atomic region. This has the advantage that the missing
terms in a truncated partial wave expansion are partly accounted for
by plane waves, which explains the rapid convergence of the partial
wave expansions.

Frequently, the question comes up, whether the transformation
Eq.~\ref{eq:transf} of the auxiliary wave functions indeed provides
the true wave function. The transformation should be considered merely
as a change of representation analogous to a coordinate transform.  If
the total energy functional is transformed consistently, its minimum
will yield an auxiliary wave function that produces a correct wave
function $|\Psi\rangle$.  

%
\section{Expectation values}
Expectation values can be obtained either from the reconstructed
true wave functions or directly from the auxiliary wave
functions 
\begin{eqnarray}
\langle A\rangle&=&\sum_{n}f_n\langle\Psi_n|A|\Psi_n\rangle
+\sum_{n=1}^{N_c}\langle\phi_n^c|A|\phi_n^c\rangle
\nonumber\\
&=&\sum_{n}f_n\langle\tilde\Psi_n|{\cal T}^\dagger A{\cal T}
|\tilde\Psi_n\rangle
+\sum_{n=1}^{N_c}\langle\phi_n^c|A|\phi_n^c\rangle
\end{eqnarray}
where $f_n$ are the occupations of the valence states and $N_c$ is the
number of core states. The first sum runs over the valence states, and
second over the core states $|\phi^c_n\rangle$.

Now we can decompose the matrix elements into their individual
contributions according to Eq.~\ref{eq:aewave}.
\begin{eqnarray}
\langle\Psi|A|\Psi\rangle&=&
\langle\tilde\Psi+\sum_R(\Psi^1_R-\tilde\Psi^1_R)|
A|\tilde\Psi+\sum_{R'}(\Psi^1_{R'}-\tilde\Psi^1_{R'})\rangle
\nonumber\\
&=&\underbrace{\langle\tilde\Psi|A|\tilde\Psi\rangle
+\sum_R\Bigl(\langle\Psi^1_R|A|\Psi^1_R\rangle
-\langle\tilde\Psi^1_R|A|\tilde\Psi^1_R\rangle\Bigr)}_{\mbox{part 1}}
\nonumber\\
&+&\underbrace{\sum_R\Bigl(
\langle \Psi^1_R-\tilde\Psi^1_R|A|\tilde\Psi-\tilde\Psi^1_R\rangle
+\langle\tilde\Psi-\tilde\Psi^1_R|A|\Psi^1_R-\tilde\Psi^1_R\rangle
\Bigr)}_{\mbox{part 2}}
\nonumber\\
&+&\underbrace{\sum_{R\neq R'}\langle \Psi^1_R
-\tilde\Psi^1_R|A|\Psi^1_{R'}-\tilde\Psi^1_{R'}\rangle}
_{\mbox{part 3}}
\label{eq:expect}
\end{eqnarray}
Only the first part of Eq.~\ref{eq:expect}, is evaluated explicitly,
while the second and third parts of Eq.~\ref{eq:expect} are neglected,
because they vanish for sufficiently local operators as long as the
partial wave expansion is converged: The function
$\Psi^1_R-\tilde\Psi^1_R$ vanishes per construction beyond some
augmentation region, because the partial waves are pairwise identical
beyond that region. The function $\tilde\Psi-\tilde\Psi^1_R$ vanishes
inside the augmentation region, if the partial wave expansion is
sufficiently converged.  In no region of space both functions
$\Psi^1_R-\tilde\Psi^1_R$ and $\tilde\Psi -\tilde\Psi^1_R$ are
simultaneously nonzero.  Similarly the functions
$\Psi^1_R-\tilde\Psi^1_R$ from different sites are never non-zero in
the same region in space. Hence, the second and third parts of
Eq.~\ref{eq:expect} vanish for operators such as the kinetic energy
$\frac{-\hbar^2}{2m_e}\mathbf{\nabla}^2$ and the real space projection
operator $|r\rangle\langle r|$, which produces the electron density.
For truly nonlocal operators the second and third parts of
Eq.~\ref{eq:expect} would have to be considered explicitly.

The expression, Eq.~\ref{eq:expect}, for the expectation value can therefore be written as
\begin{eqnarray}
\langle A\rangle&=&
\sum_{n}f_n\Bigl(\langle\tilde\Psi_n|A|\tilde\Psi_n\rangle
+\langle\Psi^1_n|A|\Psi^1_n\rangle
-\langle\tilde\Psi^1_n|A|\tilde\Psi^1_n\rangle\Bigr)
+\sum_{n=1}^{N_c}\langle\phi^c_n|A|\phi^c_n\rangle
\nonumber\\
&=&\sum_{n}f_n\langle\tilde\Psi_n|A|\tilde\Psi_n\rangle
+\sum_{n=1}^{N_c}\langle\tilde\phi^c_n|A|\tilde\phi^c_n\rangle
\nonumber\\
&+&\sum_R\Bigl(\sum_{i,j\in R}D_{i,j}\langle\phi_j|A|\phi_i\rangle
+\sum_{n\in R}^{N_{c,R}}\langle\phi^c_n|A|\phi^c_n\rangle\Bigr)
\nonumber\\
&-&\sum_R\Bigl(\sum_{i,j\in R} D_{i,j}\langle\tilde\phi_j|A|\tilde\phi_i\rangle
+\sum_{n\in R}^{N_{c,R}}\langle\tilde\phi^c_n|A|\tilde\phi^c_n\rangle\Bigr)
\end{eqnarray}
where $D_{i,j}$ is the one-center density matrix defined as
\begin{eqnarray}
D_{i,j}=\sum_n
f_n\langle\tilde\Psi_n|\tilde{p}_j\rangle
\langle\tilde{p}_i|\tilde\Psi_n\rangle
=\sum_n \langle\tilde{p}_i|\tilde\Psi_n\rangle
f_n\langle\tilde\Psi_n|\tilde{p}_j\rangle
\label{eq:1cdenmat}
\end{eqnarray}

The auxiliary core states, $|\tilde\phi^c_n\rangle$ allow to incorporate the tails of the core
wave function into the plane wave part, and therefore assure, that the
integrations of partial wave contributions cancel strictly beyond
$r_c$. They are identical to the true core states in the tails, but
are a smooth continuation inside the atomic sphere. It is not required
that the auxiliary wave functions are normalized.

For example, the electron density is given by
\begin{eqnarray}
n(\mathbf{r})&=&\tilde{n}(\mathbf{r})
+\sum_R\Bigl(n^1_R(\mathbf{r})-\tilde{n}^1_R(\mathbf{r})\Bigr)
\\
\tilde{n}(\mathbf{r})&=&\sum_n f_n \tilde\Psi^*_n(\mathbf{r})\tilde\Psi_n(\mathbf{r})+\tilde{n}_c
\nonumber\\
n^1_R(\mathbf{r})&=&\sum_{i,j\in R}
D_{i,j}\phi^*_j(\mathbf{r})\phi_i(\mathbf{r})+n_{c,R} \nonumber
\nonumber\\
\tilde{n}^1_R(\mathbf{r})&=&\sum_{i,j\in R}
D_{i,j}\tilde\phi^*_j(\mathbf{r})\tilde\phi_i(\mathbf{r})+\tilde{n}_{c,R}
\end{eqnarray}
where $n_{c,R}$ is the core density of the corresponding atom and
$\tilde{n}_{c,R}$ is the auxiliary core density that is identical to $n_{c,R}$
outside the atomic region and a smooth continuation inside.

Before we continue, let us discuss a special point: The matrix element
of a general operator with the auxiliary wave functions may be slowly
converging with the plane wave expansion, because the operator $A$ may
not be well behaved. An example for such an operator is the singular
electrostatic potential of a nucleus. This problem can be alleviated by
adding an intelligent zero: If an operator $B$ is purely localized
within an atomic region, we can use the identity between the auxiliary
wave function and its own partial wave expansion
\begin{eqnarray}
0&=&\langle\tilde\Psi_n|B|\tilde\Psi_n\rangle
-\langle\tilde\Psi_n^1|B|\tilde\Psi_n^1\rangle
\label{eq:zeroop}
\end{eqnarray}
Now we choose an operator $B$ so that it cancels the problematic
behavior of the operator $A$, but is localized in a single atomic
region.  By adding $B$ to the plane wave part and the matrix elements
with its one-center expansions, the plane wave convergence can be
improved without affecting the converged result.

\section{Total Energy} 
Like wave functions and expectation values also the total
energy can be divided into three parts.
\begin{eqnarray}
E([\tilde\Psi_n],R_i)&=&\tilde{E}+\sum_R\Bigl(E^1_R-\tilde{E}^1_R\Bigr)
\end{eqnarray}
The plane-wave part $\tilde{E}$ involves only smooth functions and is
evaluated on equi-spaced grids in real and reciprocal space.  This
part is computationally most demanding, and is similar to the
expressions in the pseudopotential approach.  
\begin{eqnarray}
\tilde{E}&=&\sum_n\langle\tilde\Psi_n|\frac{-\hbar^2}{2m_e}\nabla^2
|\tilde\Psi_n\rangle
\nonumber\\
&+&\frac{e^2}{8\pi\epsilon_0}\int d^3r\int d^3r'
\frac{[\tilde{n}(\mathbf{r})+\tilde{Z}(\mathbf{r})]
[\tilde{n}(\mathbf{r}')+\tilde{Z}(\mathbf{r}')]}{|\mathbf{r}-\mathbf{r}'|}
\nonumber\\
&+&\int d^3r \tilde{n}(\mathbf{r})\epsilon_{xc}(\mathbf{r},[\tilde{n}])
+\int d^3r \bar{v}(\mathbf{r})\tilde{n}(\mathbf{r}),
\label{eq:psetot}
\end{eqnarray}
where $\tilde{Z}({\bf r})$ is an angular dependent core-like density that 
will be described in detail below.
The remaining parts can be evaluated on radial grids in a spherical
harmonics expansion. 
The nodal structure of the wave functions can be properly described
on a logarithmic radial grid that becomes very fine near nucleus, 
\begin{eqnarray}
{E}^1_R&=&\sum_{i,j\in R} D_{i,j}
\langle\phi_j|\frac{-\hbar^2}{2m_e}\nabla^2|\phi_i\rangle
+\sum_{n\in R}^{N_{c,R}}
\langle\phi^c_n|\frac{-\hbar^2}{2m_e}\nabla^2|\phi^c_n\rangle
\nonumber\\
&+&\frac{e^2}{8\pi\epsilon_0}\int d^3r\int d^3r'
\frac{[n^1(\mathbf{r})+Z(\mathbf{r})][n^1(\mathbf{r}')+Z(\mathbf{r}')]}{|\mathbf{r}-\mathbf{r}'|}
\nonumber\\
&+&\int d^3r n^1(\mathbf{r})\epsilon_{xc}(\mathbf{r},[n^1])
\\
\tilde{E}^1_R&=&\sum_{i,j\in R} D_{i,j}
\langle\tilde\phi_j|\frac{-\hbar^2}{2m_e}\nabla^2|\tilde\phi_i\rangle
\nonumber\\
&+&\frac{e^2}{8\pi\epsilon_0}\int d^3r\int d^3r'
\frac{[\tilde{n}^1(\mathbf{r})+\tilde{Z}(\mathbf{r})][\tilde{n}^1(\mathbf{r}')+\tilde{Z}(\mathbf{r}')]}{|\mathbf{r}-\mathbf{r}'|}
\nonumber\\
&+&\int d^3r \tilde{n}^1(\mathbf{r})\epsilon_{xc}(\mathbf{r},[\tilde{n}^1])
+\int d^3r \bar{v}(\mathbf{r})\tilde{n}^1(\mathbf{r})
\label{eq:ps1etot}
\end{eqnarray}

The nuclear charge density $-eZ(\mathbf{r})$ is defined as a sum of
$\delta$-functions on the nuclear sites, $Z(\mathbf{r})=-\sum_R {\cal
Z}_R\delta(\mathbf{r}-\mathbf{R})$, with the atomic numbers ${\cal
Z}_R$.  Note that the self energy of a point charge is infinite and
must be subtracted out.

The compensation density
$\tilde{Z}(\mathbf{r})=\sum_R\tilde{Z}_R(\mathbf{r})$ is given as a
sum of angular momentum dependent Gauss functions, which have an
analytical Fourier transform.  A similar term occurs also in the
pseudopotential approach. In contrast to the norm-conserving
pseudopotential approach however, the compensation charge is
non-spherical and it is constantly adapting to the instantaneous
environment. It is constructed such that the augmentation charge
densities
\begin{eqnarray}
n^1_R(\mathbf{r})+Z_R(\mathbf{r}) -\tilde{n}^1_R(\mathbf{r})
-\tilde{Z}_R(\mathbf{r})
\end{eqnarray}
have vanishing electrostatic multi-pole moments for each atomic site.
As a result the sum of all one-center contributions from one atom does
not produce an electrostatic potential outside their own atomic
region. This is the reason that the electrostatic interaction of the
one-center parts between different sites vanish.

The compensation charge density as given here is still localized
within the atomic regions, but a technique similar to an Ewald
summation allows to replace it by a very extended charge density. Thus
we can achieve, that all functions in $\tilde{E}$ converge as fast as
the auxiliary density itself.

The potential $\bar{v}$, which occurs in
Eqs.~\ref{eq:psetot} and~\ref{eq:ps1etot} enters the total energy in the
form of a zero described in Eq.~\ref{eq:zeroop}
\begin{eqnarray}
\sum_nf_n\langle\tilde\Psi_n|
\Bigl(\bar{v}-\sum_{i,j}|\tilde{p}_i\rangle
\langle\tilde\phi_i|\bar{v}|\tilde\phi_j\rangle
\langle\tilde{p}_j|\Bigr)|\tilde\Psi_n\rangle
\end{eqnarray}
The main reason for introducing this potential is that the
self-consistent potential resulting from the plane wave part is not
necessarily optimally smooth.  The potential $\bar{v}$ allows to
influence the plane wave convergence beneficially, without changing
the converged result.  $\bar{v}$ must be localized within the augmentation region, 
where equation~\ref{eq:ps1center} holds.
%
\section{Approximations}
Once the total energy functional provided in the previous section has
been defined, everything else follows: Forces are partial derivatives
with respect to atomic positions. The potential is the derivative of
the potential energy with respect to the density, and the Hamiltonian
follows from derivatives with respect to wave functions.  The
fictitious Lagrangian approach of Car and Parrinello \cite{Car85} does
not allow any freedom in the way these derivatives are
obtained. Anything else than analytic derivatives will violate energy
conservation in a dynamical simulation. Since the expressions are
straightforward, even though rather involved, we will not discuss them
here.

All approximations are incorporated already in the total energy
functional of the PAW method. What are those approximations?
\begin{itemize}
\item Firstly we use the frozen core approximation. In principle this
approximation can be overcome.
\item The plane wave expansion for the auxiliary wave functions must
be complete. The plane wave expansion is controlled easily by
increasing the plane wave cutoff defined as $E_{PW}=\frac{1}{2}\hbar^2G_{max}^2$.
Typically we use a plane wave cutoff of 30~Ry.
\item The partial wave expansions must be converged. Typically we use
one or two partial waves per angular momentum $(\ell,m)$ and site.  It
should be noted that the partial wave expansion is not variational,
because the partial wave expansion changes the total energy functional
and not only the basis set.
\end{itemize}
We do not discuss here numerical approximations such as the choice of
the radial grid, since those are easily controlled.

We mentioned earlier that the pseudopotential approach can be derived
as a well defined approximation from the PAW method:
The augmentation part $\Delta E=E^1-\tilde{E}^1$ is a functional of
the one-center density matrix $D_{i,j}$ defined in
Eq.~\ref{eq:1cdenmat}. The pseudopotential approach can be recovered
if we truncate a Taylor expansion of $\Delta E$ about the atomic
density matrix after the linear term.  The term linear to $D_{i,j}$ is
the energy related to the nonlocal pseudopotential.
\begin{eqnarray}
\Delta E(D_{i,j})&=&\Delta E(D_{i,j}^{at})
+\sum_{i,j}(D_{i,j}-D^{at}_{i,j})\frac{\partial \Delta E}{\partial D_{i,j}}
+O(D_{i,j}-D^{at}_{i,j})^2
\nonumber\\
&=&E_{self}+\sum_n f_n\langle\tilde\Psi_n|v_{nl}|\tilde\Psi_n\rangle
+O(D_{i,j}-D^{at}_{i,j})^2
\end{eqnarray}
Thus we can look at the PAW method also as a pseudopotential method
with a pseudopotential that adapts to the instantaneous electronic
environment, because the explicit nonlinear dependence of the total
energy on the one-center density matrix is properly taken into
account.

What are the main advantages of the PAW method compared to the
pseudopotential approach?  

Firstly all errors can be systematically controlled so that there are
no transferability errors.  As shown by Watson \cite{Watson98} and
Kresse \cite{Kresse99}, most pseudopotentials fail for high spin atoms
such as Cr. While it is probably true that pseudopotentials can be
constructed that cope even with this situation, a failure can not be
known beforehand, so that some empiricism remains in practice: A
pseudopotential constructed from an isolated atom is not guaranteed to
be accurate for a molecule.  In contrast, the converged results of the
PAW method do not depend on a reference system such as an isolated
atom, because it uses the full density and potential.

The PAW method provides access to the full charge and spin density,
which is relevant for hyperfine parameters. Hyperfine parameters are
sensitive probes of the electron density near the nucleus. In many
situations they are the only information available that allows to
deduce atomic structure and chemical environment of an atom.  There
are reconstruction techniques for the pseudopotential approach, which
however, are poor man's versions \cite{Hyperfine} of the PAW method.

The plane wave convergence is more rapid than in norm-conserving
pseudopotentials and should in principle be equivalent to that of
ultra-soft pseudo\-potentials \cite{Vanderbilt90}. Compared to the
ultra-soft pseudo\-potentials, however, the PAW method has the advantage
that the total energy expression is less complex and therefore is
expected to be more efficient.

The construction of pseudopotentials requires to determine a number of
parameters. As they influence the results, their choice is critical.
Also the PAW methods provides some flexibility in the choice of
auxiliary partial waves. However, this choice does not influence the
converged results.

\section{Recent Developments}

Since the first implementation of the PAW method in the CP-PAW
code, a number of groups have adopted the PAW method.  The
second implementation was done by the group of
Holzwarth \cite{Holzwarth97}. The resulting PWPAW code is freely
available \cite{Holzwarth01}.  This code is also used as a basis for
the PAW implementation in the AbInit project \cite{Abinit}. An
independent PAW code has been developed by Valiev and
Weare \cite{Valiev99}. Recently the PAW method has been implemented
into the VASP code \cite{Kresse99}. The PAW method has also been
implemented by W. Kromen into the ESTCoMPP code of Bl\"ugel and
Schr\"oder \cite{Kromen}.

Another branch of methods uses the reconstruction of the PAW
method, without taking into account the full wave functions in the
self-consistency. Following chemist notation this approach could be
termed ``post-pseudopotential PAW''. This development began with the
evaluation for hyperfine parameters from a pseudopotential calculation
using the PAW reconstruction operator \cite{Hyperfine} and is now used
in the pseudopotenital approach to calculate properties that 
require the correct wave functions

The implementation by Kresse and Joubert \cite{Kresse99} has been
particularly useful as they had an implementation of PAW in the same
code as the ultra-soft pseudo\-potentials, so that they could critically
compare the two approaches with each other and LAPW
calculations. Their conclusion is that both methods compare well in
most cases, but they found that magnetic energies 
are seriously -- by a factor two -- in error in the pseudo\-potential approach,
while the results of the PAW method were in line with other
all-electron calculations using the linear augmented plane wave
method. As a short note, Kresse and Joubert incorrectly claim that their
implementation is superior as it includes a term that is analogous to
the non-linear core correction of pseudo\-potentials \cite{NLCC}: this
term however is already included in the original version in the 
form of the pseudized core density.

Several extensions of the PAW have been done in the recent years:
For applications in chemistry truly isolated systems are often of
great interest. As any plane-wave based method introduces periodic
images, the electrostatic interaction between these images can cause
serious errors. The problem has been solved by mapping the charge
density onto a point charge model, so that the electrostatic
interaction could be subtracted out in a self-consistent
manner \cite{decoupling}. In order to include the influence of the
environment, the latter was simulated by simpler force fields using
the molecular-mechanics-quantum-mechanics (QM-MM) approach \cite{QMMM}.

In order to overcome the limitations of the density functional theory
several extensions have been performed. Bengone \cite{Bengone00}
implemented the LDA+U approach \cite{LDA+U} into the CP-PAW code.  Soon
after this, Arnaud \cite{Arnaud2000} accomplished the implementation of the
GW approximation into the CP-PAW code.  The VASP-version of
PAW \cite{Kresse00} and the CP-PAW code have now been extended to
include a noncollinear description of the magnetic moments. In a
non-collinear description the Schr\"odinger equation is replaced by
the Pauli equation with two-component spinor wave functions

The PAW method has proven useful to evaluate electric field
gradients \cite{EFG} and magnetic hyperfine parameters with high
accuracy \cite{SiO2}. Invaluable will be the prediction of NMR chemical
shifts using the GIPAW method of Pickard and Mauri \cite{Mauri2001},
which is based on their earlier work \cite{Mauri96}. While the GIPAW
is implemented in a post-pseudo\-potential manner, the extension to a
self-consistent PAW calculation should be straightforward.  An
post-pseudo\-potential approach has also been used to evaluate core
level spectra \cite{Pickard97} and momentum matrix
elements \cite{Kageshima}.

\section{Acknowledgment}

We are grateful for carefully reading the manuscript to Dr.\,J.\,Noffke, 
Dr.\,R.\,Schmidt and P.\,Walther and to Prof.\,K.\,Schwarz for his 
continuous support. This work has benefited from the collaborations 
within the ESF Programme on 'Electronic Structure Calculations for 
Elucidating the Complex Atomistic Behavior of Solids and Surfaces'.  

\bibliographystyle{plain}

%
\end{document}